# Ultrathin plasma polymer passivation of perovskite solar cells for improved stability and reproducibility

*Jose M. Obrero-Perez, Lidia Contreras-Bernal+, Fernando Nuñez-Galvez, Javier Castillo-Seoane, Karen Valadez-Villalobos, Francisco J. Aparicio, Juan A. Anta, Ana Borras, Juan R. Sanchez-Valencia, and Angel Barranco+*

J. Obrero-Perez, Dr. L. Contreras-Bernal, F. Nuñez-Galvez, J. Castillo-Seoane, Dr. F. J. Aparicio, Dr. A. Borras, Dr. J. R. Sanchez-Valencia and Dr. A. Barranco
Instituto de Ciencia de Materiales de Sevilla (CSIC-Universidad de Sevilla)
C/Americo Vespucio 49, E-41092 Seville, Spain
E-mail: lidia.contreras@icmse.csic.es; angel.barranco@csic.es

Dr. K. Valadez-Villalobos and J. A. Anta
Área de Química Física
Universidad Pablo de Olavide E-41013 Seville, Spain

Keywords: perovskite solar cells, plasma and vacuum depositions, ultrathin polymers, stability, passivation interfaces.

Despite the youthfulness of hybrid halide perovskite solar cells, their efficiencies are currently comparable to commercial silicon and have surpassed quantum-dots solar cells. Yet, the scalability of these devices is a challenge due to their low reproducibility and stability under environmental conditions. However, the methods reported to date to tackle such issues recurrently involve the use of solvent methods that would further complicate their transfer to industry. Herein we present a reliable alternative relaying in the implementation of an ultrathin plasma polymer as passivation interface between the electron transport material and the hybrid perovskite layer.  Such nanoengineering interface provides solar devices with increased long-term stability under ambient conditions. Thus, without consideringr any additional encapsulation step, the cells retain more than 80 % of their efficiency after being exposed to the ambient atmosphere for more than 1000 h. Moreover, this plasma polymer passivation strategy significantly improves the coverage of the mesoporous scaffold by the perovskite layer, providing the solar cells with enhanced performance as well as improved reproducibility.

1. Introduction

Under the scenario of the climate crisis caused by human actions,[1,2] the photovoltaic field has undergone rapid progress in the last few years due to the development of solar devices based on hybrid metal halide perovskite materials.[3] To date, the most highly efficient perovskite solar cells (PSCs) are been achieved by mesoscopic architecture using mesoporous $TiO_2$ (*m*-$TiO_2$) as electron transport layer (ETL). The mesoporous scaffold is usually doped with hygroscopic compounds as lithium salt to enhance its electron mobility.[4–7] Although Li-treatment reduces the recombination loss at the ETL, it also causes greater instability of the solar devices against ambient moisture as well as low reproducibility of their photovoltaic parameters.[8,9] Indeed, one of the current handicaps of PSCs is their sensitivity to environmental conditions since, among other aspects, the perovskite material degrades rapidly upon exposure to light, humidity, oxygen, or/and high temperatures.[10] Besides that, this lack of stability to the air atmosphere imposes the use of a dry atmosphere glove-box which hampers the large-scale production of these solar devices.[11–15] In this context, researchers have focused on finding interface passivation alternatives that are not detrimental to the performance of the devices. So far, the materials employed to passivate interfaces have been 2D perovskites, metal oxides compounds, or insulating organic materials. These reported approaches typically use solutions methods, however, no alternative vacuum processes scalable to industrial manufacturing have been explored.[16–18]

In this work, we propose the use of ultrathin plasma polymers as passivating material for the ETL/perovskite interface. These plasma polymeric films have already been successfully tested as encapsulating material for protecting PSC against water and moisture and also as conformal protective layers for organic supported nanostructures.[19,20] The films are grown in a single step by remote plasma assisted vacuum deposition (RPAVD) of adamantane powder.[20,21] This molecule is a cage-like hydrocarbon consisting of $sp^3$ C-C bonds with the most diamondoid structure, which is vacuum sublimated in presence of low-power remote microwave plasma. Applying this method, adamantane-based ultrathin plasma polymers were deposited at room temperature on the mesoporous titanium oxide (*m*-$TiO_2$/ADA) without any required pretreatment. Their microstructural characteristics have been observed by scanning electron microscopy (SEM). The plasma polymer resulting from this process is highly conformal, homogenous, compact, and chemically inert, as we have already shown in our previous works.[19,20] In addition, the ADA films are partly hydrophobic, fully stable in air, and thermally stable up to 250 ºC in air. The photovoltaic behavior of the resulting solar cells has been analyzed under solar illumination conditions and the electronic effect of ADA polymeric passivation interface has been studied by electrochemical impedance. The long-term stability under ambient conditions has been analyzed and compared with those without ADA plasma polymeric passivation. It is worth highlighting, that the reported plasma polymerization fabrication technique is solventless, environmentally friendly, and fully compatible with large-scale fabrication.[22] To our best knowledge, this is the first reported work in which ultrathin plasma polymers are used at ETL/perovskite interface for PSC.

2. Results and discussion

*2.1. Adamantane plasma polymers on mesoporous $TiO_2$ electrode*

A set of adamantane plasma polymer layers with different thicknesses has been tested as a passivation layer at m-$TiO_2$/perovskite interface for conventional *n-i-p* PSCs. These ADA films were deposited on top of the mesoporous $TiO_2$ layer using the RPAVD reported in previous work.[19,20] A simplified schematic of the deposition process is shown in **Figure 1**a. In particular, ADA nanocomposite films of ca. 1nm, 5nm and 15 nm (hereinafter referred to as ADA-1nm, ADA-5nm and ADA-15nm, respectively) have been prepared. The thickness of every ADA layer has been in-situ monitored during deposition using a Quartz Crystal Monitor (see experimental section). In addition, the thickness was verified by Variable-Angle Ellipsometric Spectroscopy (VASE), whose spectra were fitted assuming a Cauchy model. The optical constant of the model, as well as the thickness, are shown in Table S1 in Supporting Information. The cross-section SEM image of the thicker ADA film (m-$TiO_2$/ADA-15nm) displays the characteristic morphology of a plasma polymer, forming a compact and conformal layer on top of the mesoporous titania (see Figure S1 and 1-a).

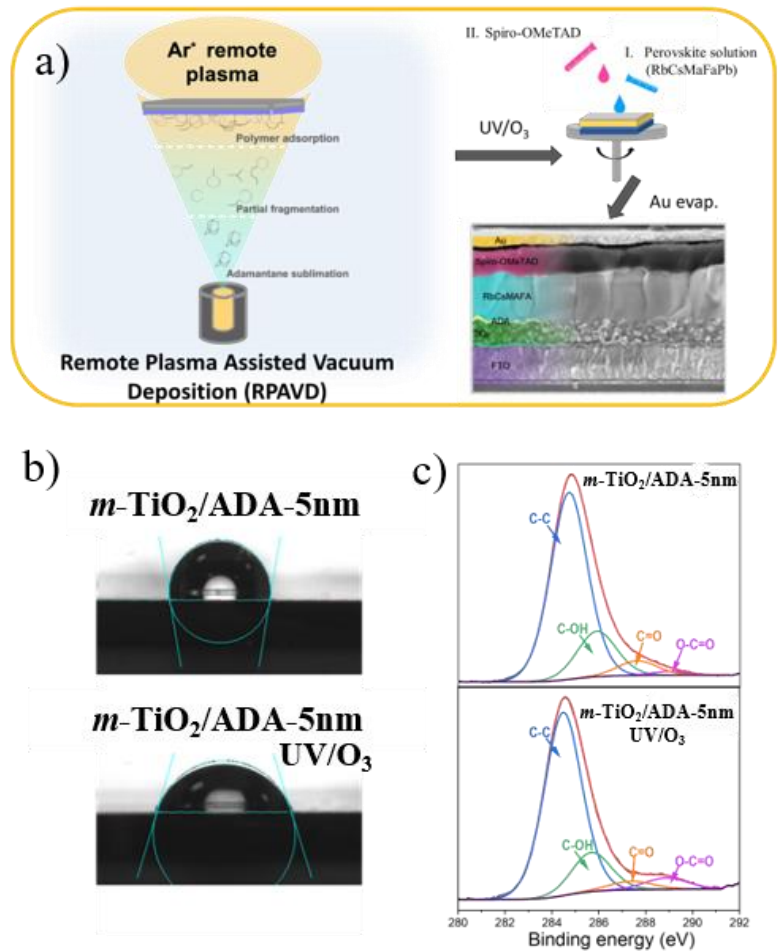

**Figure 1**. a) Schematic of the solar cell preparation process for devices based on *m*-TiO₂/ADA electrodes. A cross-sectional SEM image of a solar device with 15 nm of plasma adamantane polymer is included. b) Static water contact angle evolution after the UV/O$_3$ treatment. b) C 1s XPS peaks after and before the UV/O$_3$ surface activation treatment of t m-TiO₂/ADA-5nm surface.

*2.2. m -TiO₂/ADA as electron selective contact in perovskite solar cells*

Figure 1a shows a schematic diagram of the device manufacturing process for solar devices based on *m*-TiO₂/ADA electrodes: using these electrodes as the electron transport layer, the configuration of the solar device was completed with RbCsMAFA perovskite [((FAPbI$_3$)$_{83}$(MAPbBr$_3$)$_{17}$ + 5% CsI)+ 5% RbI] as the active layer and Spiro-OMeTAD as the hole transport layer (HTL) (see the experimental section for further details).[8] *m*-TiO₂ electrodes without ADA films were used as references samples (ADA-0nm). Since deposited ADA films are partly hydrophobic, the perovskite solution does not wet the *m*-TiO₂/ADA samples. Thus, to achieve a full and homogeneous distribution of the perovskite film it was necessary to apply a short treatment with UV ozone cleaner (hereafter UV/O$_3$ treatment) for 5 min to decrease the contact angle of the *m*-TiO$_2$/ADA surface (see Figure 1b). On these partially hydrophilic surfaces, the perovskite is optimally distributed. The UV/O$_3$ treatment hardly affected the surface composition of the polymer layer as seen by X-Ray Photoelectron Spectroscopy (XPS) analysis. In particular, only the C1s XPS peak revealed a slight increase due to the contribution of carboxyl groups shown at higher binding energies (Figure 1c). It is worth highlighting that the UV/O$_3$ treatment does not produce a significant decrease of the ADA thickness, as it was verified by VASE measurements (see Table S1).

Cross-sectional SEM images of each of the studied cell configurations are shown in **Figure 2** In all cases, the devices exhibit a conventional profile in which the perovskite layers present large and well-defined grains, as well as a thickness of around 600 nm (∼ 150 nm of *m*-TiO$_2$). We found that the mesoporous particles of TiO$_2$ were less visible in the case of ADA film was incorporated into the architecture. In other words, the presence of the ADA interlayer induces a better coating of the mesoporous scaffold by the perovskite material. In particular, the greatest overlapping at the *m*-TiO$_2$/perovskite interface came from ADA-5nm sample. This indicates that the polymeric interface is beneficial for improving the contact between the ETL and the perovskite. This key feature would lead to higher charge injection and transport, and hence to a better performance of the whole device.

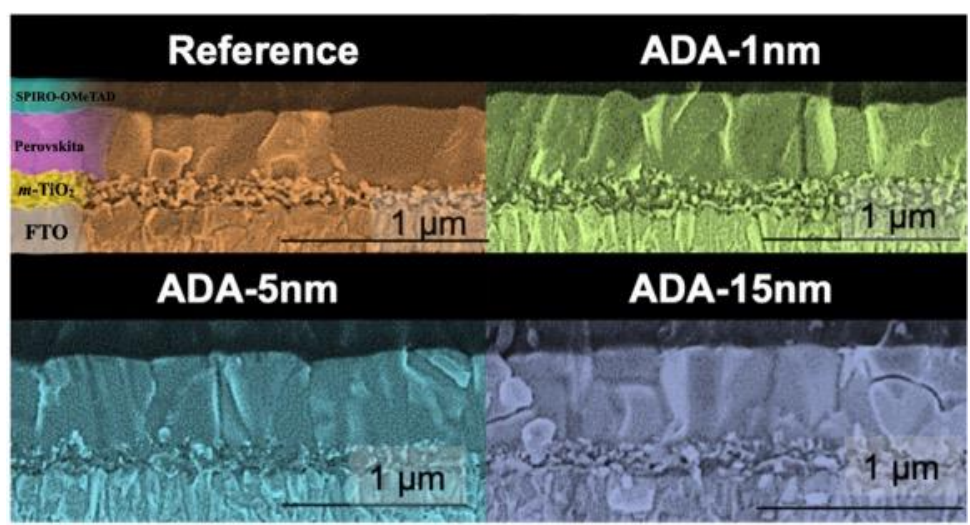

**Figure 2**. Cross-sectional SEM micrographs of complete perovskite solar cells with and without ADA polymer at the *m*-TiO$_2$/perovskite interface

**Figure 3**a) displays the current density-voltage curves (*J-V* curves) of the solar devices for the different studied ADA cells and references without ADA film measured under 1sun – AM 1.5G illumination. It is observed that the record efficiency of 19.2% is obtained for ADA-5nm devices compared to 18.2 % for ADA-0nm samples. As it can be seen in Figure 3b-e, the ADA interlayer increase not only the power conversion efficiency (PCE) (Figure 2e), but also the other photovoltaic parameters of the device: open-circuit photovoltage ($V_{OC}$) (b), Fill Factor (c) and short-circuit photocurrent density ($J_{SC}$) (d). A significant improvement in the reproducibility of ADA samples is also observed through testing 20 samples of each configuration (Figure 3b-e and Table S2). It is worthy to mention that this statistical analysis is crucial to unravel the effect of the adamantane-based plasma polymer as an interlayer due to the reported low reproducibility of PSCs. As a result of the statistical analysis, a significant increase of $V_{OC}$ was found for ADA-5nm and ADA-15nm, *i.e.*, for samples with a polymer thickness greater than or equal to 5 nm (Figure 3b). In particular, the highest value was obtained for ADA-5nm that showed a $V_{OC}$ record of 1.15 V. In the case of $J_{SC}$, significant differences were found only for ADA layers above 15 nm (Figure 3d). ADA-15nm showed the highest $J_{SC}$ data of 23.1 mA cm$^{-2}$ versus the 22.5 mA cm$^{-2}$ measured for the reference sample. On the other hand, the highest impact of ADA polymer was observed in the FF parameter (Figure 3c). The best values of FF were achieved for the thickest ADA layers, with ADA-5nm having a record value of 79 %. It is also important to stress herein that such a high fill factor is remarkable for $TiO_2$ ETL without Li-ions doping. In this context, the highest PCEs were obtained for ADA-5nm samples (Figure 3e and Table S2).

In addition, the ADA polymer acts as a passivation interlayer that impacts on the hysteresis of *J-V* curves as well (Figure 3f and S3). The addition of a thin layer of the adamantane polymer at the interface reduces the hysteresis index, HI,[23] by more than half (*i.e.*, an average HI ~ 0.13 and ~ 0.05 for reference and ADA-5nm, respectively).

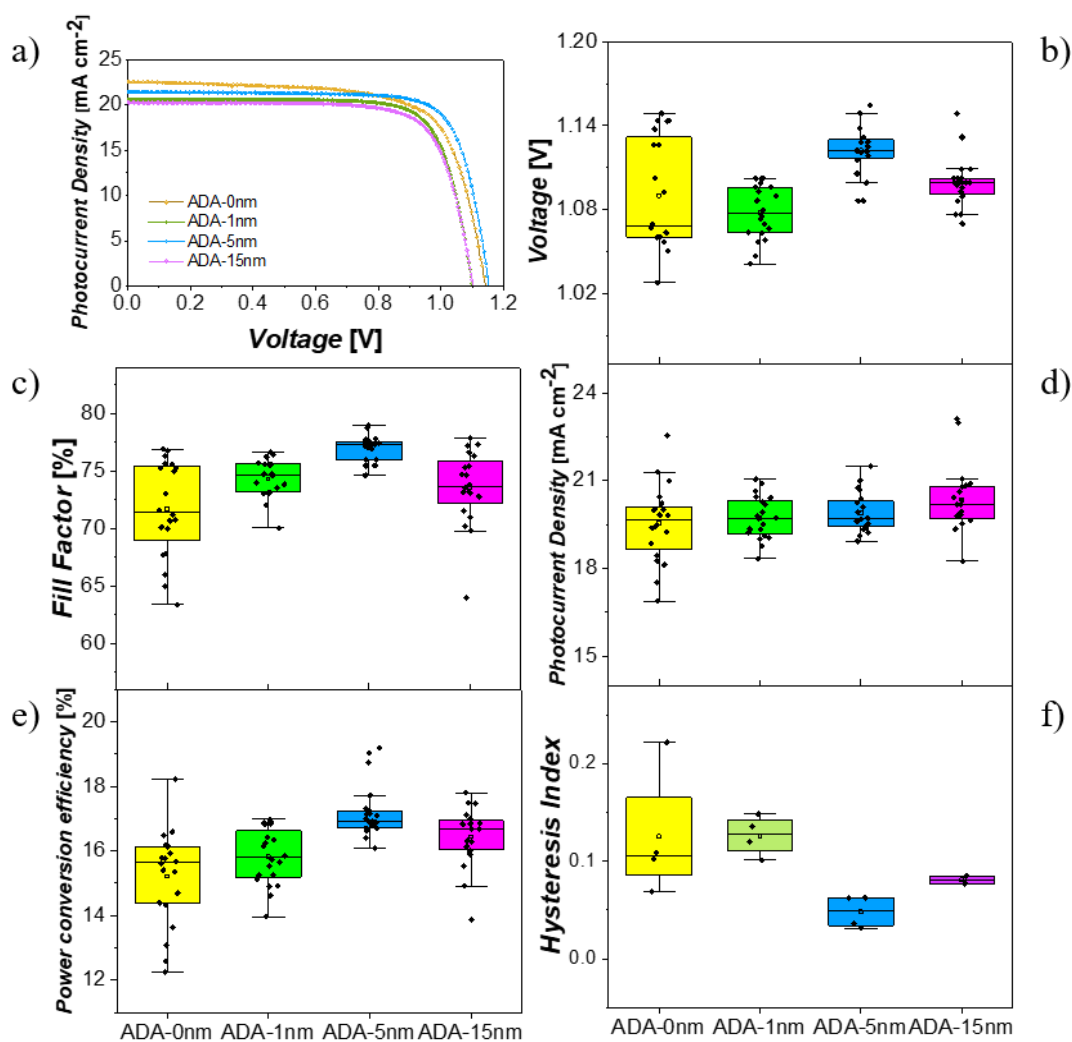

**Figure 3.** a) current-density curves of the record cells of every type (ADA-0nm, ADA-1nm, ADA-5nm and ADA-15nm). b-e) photovoltaic parameters statistics obtained from 20 different devices for each configuration using analysis of variance. These data were recorded in reverse scans: b) open-circuit potential; c) Fill factor; d) photocurrent density; e) power conversion efficiency. f) Hysteresis index (HI) statistics obtained from 4 different devices for each configuration (see the experimental section for more details).[23] The curves were measured under 1sun- AM 1.5G illumination using an illumination mask of 0.14 cm$^2$.

To cast some light on this effect, electrochemical impedance spectroscopy (IS) measurements at open-circuit voltage were carried out, as shown in **Figure 4**. This technique allows discerning the electronic processes that happen at different frequencies.[24,25] The impedance spectrum for a PSC is characterized by the presence of two signals: one at high frequency (*HF*, $10^6$-$10^3$ Hz) and the other at low frequency (*LF*, < 10 Hz). It is generally accepted that electronic transport and recombination processes affect mostly the *HF* signal while the low one is related to ionic accumulation and migration.[23,26–28] For this reason, we focus on the study of the first region. Well-defined arcs were observed in the *HF* region determined from the Nyquist plots from all samples as is typically the case for PSC (Figure 4a). For samples ADA-5nm and ADA-15nm, a lower onset of the Nyquist plots at high frequencies was found, indicating a lower series resistance (*Rs*) for these devices. This can be attributed to a better electronic transfer across the TiO$_2$/perovskite interface, consistent with the improved perovskite coverage mentioned above. IS plots were fitted by means of a simplified Voigt circuit (shown in Fig. 4a)) to extract the recombination resistance and geometrical capacitance associated with the high-frequency signal ($R_{HF}$ and $C_{HF}$, respectively). We found that the R$_{HF}$ values varied exponentially with the open-circuit potential from increased illumination light intensity, as predicted by the following equation[24,29]

$$R_{HF} = \left(\frac{\partial J_{rec}}{\partial V}\right)^{-1} = R_{00} \exp\left(-\frac{\beta qV}{k_B T}\right) \quad (1)$$

Where $J_{rec}$ is the recombination current, V is the photovoltage, $R_{00}$ is the resistance at zero potential, $\beta$ is the recombination parameter, $k_B$ is the Boltzman constant and T is the absolute temperature. Figure 4b) shows $R_{HF}$ plot from which no significant change is inferred in the recombination resistance rates. However slight changes in slope ($\beta$) were observed (see Table S3). According to previous work, the change in the $\beta$ parameter suggests a different main recombination mechanism inside of the perovskite solar cell, although some ionic effects cannot be ruled out.[25,30] On the other hand, significant changes in high-frequency capacitance are also not observed, as shown Figure 5a).

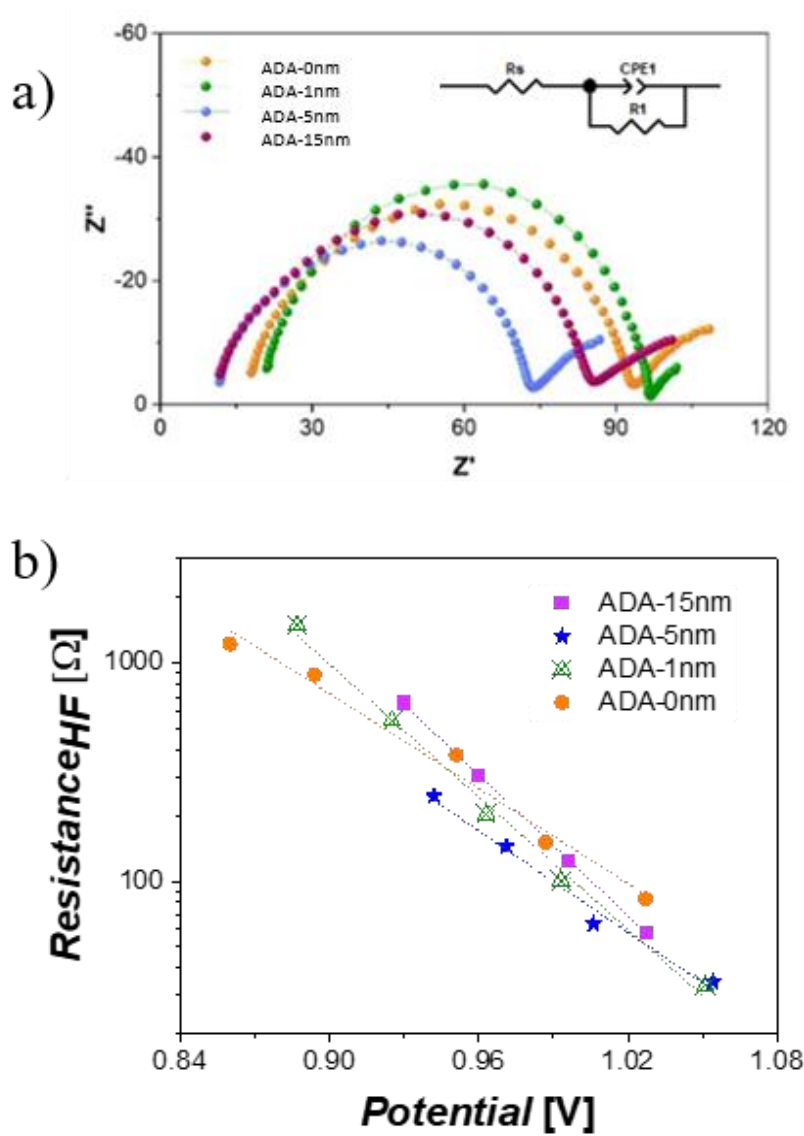

**Figure 4**. a) Nyquist plots obtained by impedance spectroscopy for ADA and reference samples (ADA-0nm). These spectra were recorded at open circuit potential, with an applied voltage of 1.008 V using red light (LED) as the illumination source. b) Resistance element extracted from Nyquist plots using simple Voight equivalent circuit fitting restricted to the high-frequency region.

Moreover, IS analyses were carried out to unravel the effect of the ultrathin ADA plasma-polymer on the electronic properties of the m-TiO$_2$ layer. For this purpose, dye-sensitized solar cells (DSSCs) based on thicker m-TiO$_2$/ADA electrodes were prepared, i.e., ADA-5nm and ADA-15nm, and ADA-0nm as reference (See Experimental section for further details). It has been widely reported that a change in the chemical capacitance (Cμ) of m-TiO2 indicates a change in the conduction band edge of the semiconductor.[31,32] $C_\mu$ from our samples did not show a clear change between ADA and reference devices (Figure 5b). That is, the band parameters of *m*-TiO$_2$ were not affected by the addition of ADA plasma polymer on the mesoporous scaffold.

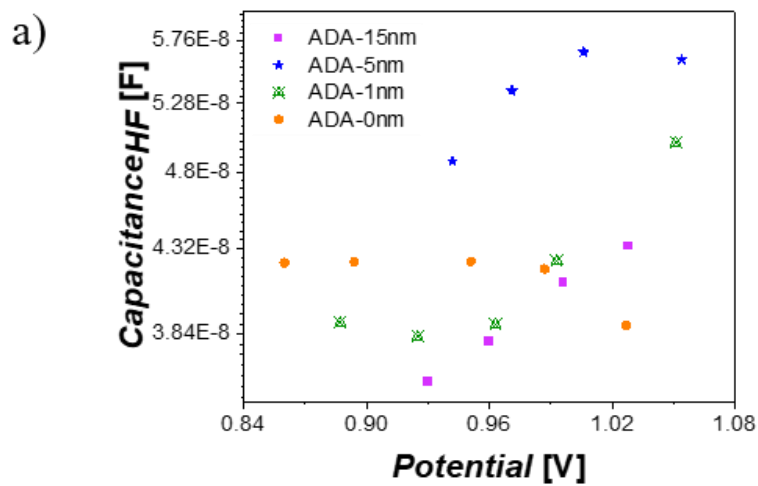

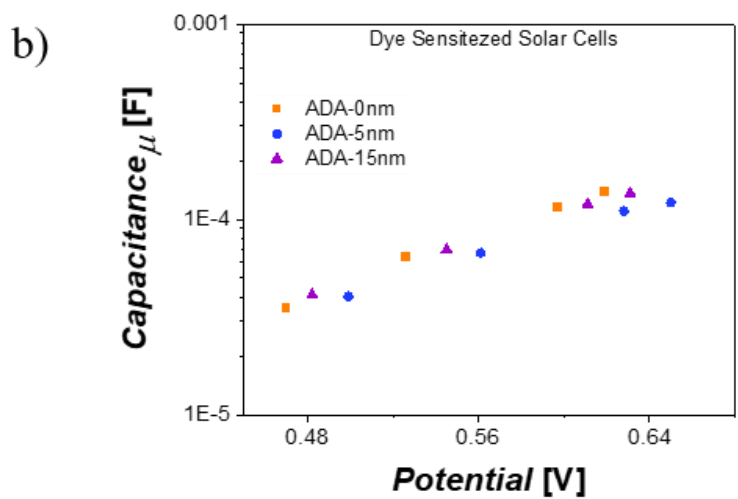

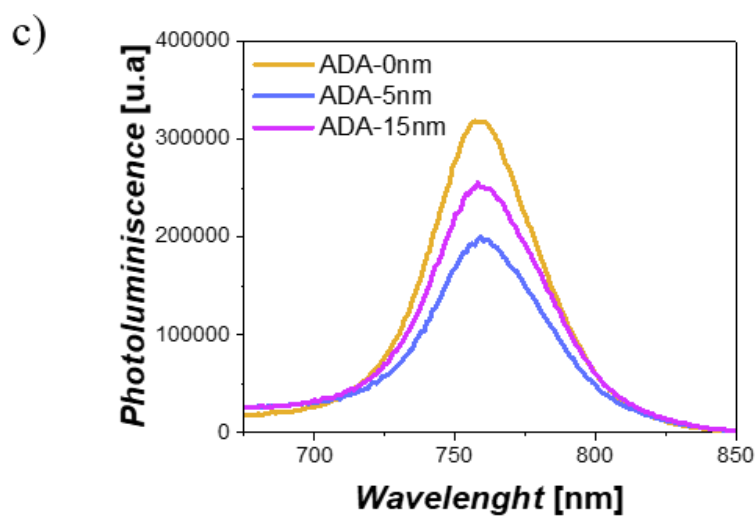

**Figure 5.** a) Capacitance element extracted from Nyquist plots using simple Voight equivalent circuit fitting restricted to the high-frequency region. b) Chemical capacitance data extracted from impedance spectroscopy measurements for dye sensitized solar cells based on ADA samples. a-b) The measurement was carried out at open circuit potential using a red LED as the illumination source. c) Steady-state photoluminescence spectra of perovskite deposited on mesoporous $TiO_2$ layer passivated ADA polymer. An excitation wavelength of 450 nm was used.

**Figure 5**c) shows the photoluminescence (PL) spectra of perovskite deposited on *m*-$TiO_2$ layer passivated with ADA polymers (reference ADA-0nm is also added). For a better comparison of intensity peak, PL data has been normalized to the maximum absorbance (Figure S2). In this way, it is possible to analyze the impact of the ADA polymer at the interlayer on both electron injection and light-harvesting. [33] The figure shows a higher PL peak intensity for the reference than ADA samples. The lowest PL signal is observed for ADA-1nm. Bearing in mind that the PL signal arises from the radiative recombination inside the perovskite film, a decreased PL intensity means faster electron injection from the perovskite layer into the $TiO_2$ layer. Therefore, a better electron transfer across the *m*-$TiO_2$/perovskite interface is obtained when ADA was incorporated into the structure, indicating that the ADA interlayer has a kinetic rather than a thermodynamic effect on PSCs.

2.3. Cell stability

In previous work, ADA films have been used as encapsulation material (with a significantly higher thickness) for solar cells to increase cell stability and resistance to moisture and water.[19] This plasma-polymer is very stable against chemical and environmental agents.[20] Here, we have investigated the long-term stability of the cells with ADA polymers at the *m*-$TiO_2$/perovskite interface under dark conditions (test I, **Figure 6** a) and lighting cycles (test II, b) both under ambient conditions (∼ 50% relative humidity and ∼ 23 ˚C). Firstly, samples were stored in air in the dark for almost 300 h. Under test I, ADA-5nm kept its initial PCE value until the end of the experiment while ADA-0nm lost 15% of PCE after 150 h of storage (Figure 6 a). Secondly, we also tested the stability of the devices under indoor lighting cycles for 9 hours per day. Figure 6b shows the normalized PCE as a function of time for the perovskite solar cells under test II conditions for more than 1500 h. Bearing in mind that our samples were not encapsulated, ADA-5nm showed good stability after more than 1000 h showing a retaining of 80 % of the original PCE. However, ADA-0nm was significantly damaged after 1000 h of storage. Figure 6c shows the appearance of the samples under test II conditions for more than 2000 h. It is clearly noticeable how the ADA-5nm sample shows no obvious signs of degradation while the reference sample appears completely degraded.

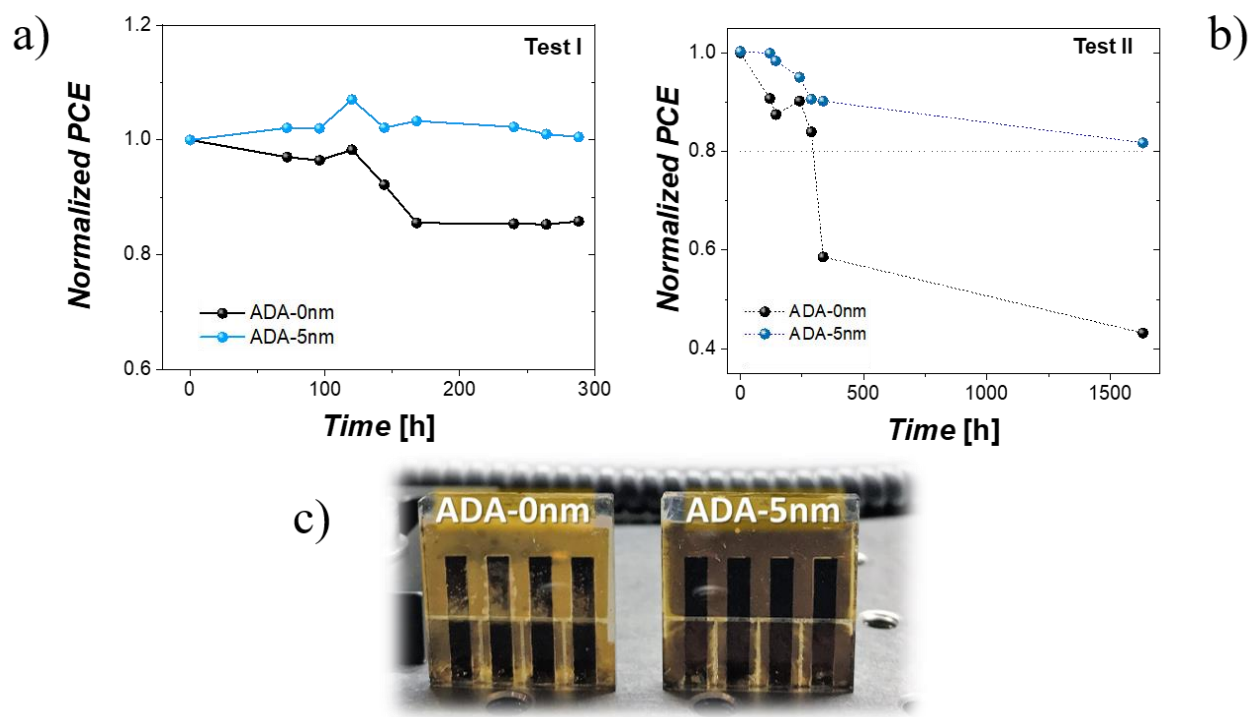

**Figure 6**. Normalized power conversion efficiency as a function of time measured under 1 sun – AM1.5G illumination in the reverse scan using a scan rate of 100 mW s$^{-1}$. The measurements were carried out at 50 % relative humidity and 23 ºC for samples stored under dark conditions (a) and for samples kept under indoor lighting cycles for 9 hours per day (b). c) Photograph of the samples after more than 2.000 hours under test II conditions.

## 3. Conclusions

The ultrathin adamantane plasma-polymeric films at *m*-TiO$_2$/perovskite interface of solar cells devices have been shown to improve significantly the reproducibility of the photovoltaic parameters. The presence of this polymeric passivation layer improves also most of the PV parameters, where it can be highlighted the increase of the fill factor. The beneficial effect on FF seems to come from the drop of series resistance observed by IS and the improvement of electronic injection found in the PL measurements. This could be due to the improved coverage of the mesoporous titanium particles by the perovskite observed in the SEM analysis of the devices. Besides that, the IS analysis of DSSC based on *m*-TiO$_2$/ADA electrodes proves that the ADA treatment did not alter the electronic properties of the TiO$_2$ electrodes. From the analysis of our results, we can establish an optimum plasma-polymer thickness of 5 nm to maximize the performance. Finally, the ADA-5nm samples show a significant stability increase under environmental conditions in comparison with the reference devices. ADA-5nm samples, measured without encapsulation at ambient conditions, retained 80 % of their initial efficiency in 1000 h.

Our results demonstrate the reliability of the plasma technique for interlayer passivation. The polymer fabrication is carried out by a solvent-free process that can be applied to large areas and/or simultaneous deposition of multiple electrodes and can be easily scaled up to industrial fabrication which could help to bring the perovskite solar cells one step closer to industry.

**Supporting Information**

Experimental section and additional characterizations.

**Acknowledgments**

The authors thank the projects PID2019-110430GB-C21, PID2019-110430GB-C22 and PID2019-109603RA-I00 funded by MCIN/AEI/10.13039/501100011033 and by "ERDF (FEDER) A way of making Europe", by the "European Union". They also thank the Consejería de Economía, Conocimiento, Empresas y Universidad de la Junta de Andalucía (PAIDI-2020 through projects US-1263142, US-1381045, US-1381057, P18-RT-3480), and the EU through cohesion fund and FEDER 2014–2020 programs for financial support. C.L.S. and J.S.-V. thank the University of Seville through the VI PPIT-US. J.S.-V.acknowledges the "Ramon y Cajal" and L.C.-B the "Juan de la Cierva" programs funded by MCIN/AEI/ 10.13039/501100011033. F.J.A. also thanks the EMERGIA Junta de Andalucía program. The projects leading to this article have received funding from the EU H2020 program under the grant agreements 851929 (ERC Starting Grant 3DScavengers). J.A.A. thanks MCIN/AEI/ 10.13039/501100011033 for SCALEUP SOLAR-ERA.net project PCI2019-111839-2.

**References**

[1] N. Oreskes, *Science* **2004**, *306*, 1686.
[2] P. C. Jain, *Renewable Energy* **1993**, *3*, 403.
[3] T. C. Sum, N. Mathews, *Energy Environ. Sci.* **2014**, *7*, 2518.
[4] E. Guillén, F. J. Ramos, J. A. Anta, S. Ahmad, *The Journal of Physical Chemistry C* **2014**, *118*, 22913.
[5] F. Giordano, A. Abate, J. P. Correa Baena, M. Saliba, T. Matsui, S. H. Im, S. M. Zakeeruddin,

M. K. Nazeeruddin, A. Hagfeldt, M. Graetzel, *Nature Communications* **2016**, *7*, 10379.
[6] A. D. Jodlowski, C. Roldán-Carmona, G. Grancini, M. Salado, M. Ralaiarisoa, S. Ahmad, N. Koch, L. Camacho, G. de Miguel, M. K. Nazeeruddin, *Nature Energy* **2017**, *2*, 972.
[7] M. Saliba, T. Matsui, K. Domanski, J.-Y. Seo, A. Ummadisingu, S. M. Zakeeruddin, J.-P. Correa-Baena, W. R. Tress, A. Abate, A. Hagfeldt, M. Gratzel, *Science* **2016**, *354*, 206.
[8] M. Saliba, J.-P. Correa-Baena, C. M. Wolff, M. Stolterfoht, N. Phung, S. Albrecht, D. Neher, A. Abate, *Chem. Mater.* **2018**, *30*, 4193.
[9] J. Chung, S. S. Shin, G. Kim, N. J. Jeon, T.-Y. Yang, J. H. Noh, J. Seo, *Joule* **2019**, *3*, 1977.
[10] B. Li, Y. Li, C. Zheng, D. Gao, W. Huang, *RSC Advances* **2016**, *6*, 38079.
[11] S. F. Lux, L. Terborg, O. Hachmöller, T. Placke, H.-W. Meyer, S. Passerini, M. Winter, S. Nowak, *Journal of The Electrochemical Society* **2013**, *160*, A1694.
[12] L. Qiu, L. K. Ono, Y. Qi, *Materials Today Energy* **2018**, *7*, 169.
[13] A. Abate, J.-P. Correa-Baena, M. Saliba, M. S. Su'ait, F. Bella, *Chemistry - A European Journal* **2018**, *24*, 3083.
[14] M. Salado, L. Contreras-Bernal, L. Caliò, A. Todinova, C. López-Santos, S. Ahmad, A. Borras, J. Idígoras, J. A. Anta, *Journal of Materials Chemistry A* **2017**, *5*, 10917.
[15] L. Contreras-Bernal, C. Aranda, M. Valles-Pelarda, T. T. Ngo, S. Ramos-Terrón, J. J. Gallardo, J. Navas, A. Guerrero, I. Mora-Seró, J. Idígoras, J. A. Anta, *The Journal of Physical Chemistry C* **2018**, *122*, 5341.
[16] A. A. Sutanto, P. Caprioglio, N. Drigo, Y. J. Hofstetter, I. Garcia-Benito, V. I. E. Queloz, D. Neher, M. K. Nazeeruddin, M. Stolterfoht, Y. Vaynzof, G. Grancini, *Chem* **2021**, *7*, 1903.
[17] L. Zuo, H. Guo, D. W. deQuilettes, S. Jariwala, N. D. Marco, S. Dong, R. DeBlock, D. S. Ginger, B. Dunn, M. Wang, Y. Yang, *Science Advances* **2017**.
[18] Q. Wang, Q. Dong, T. Li, A. Gruverman, J. Huang, *Advanced Materials* **2016**, *28*, 6734.
[19] J. Idígoras, F. J. Aparicio, L. Contreras-Bernal, S. Ramos-Terrón, M. Alcaire, J. R. Sánchez-Valencia, A. Borras, Á. Barranco, J. A. Anta, *ACS Applied Materials & Interfaces* **2018**, *10*, 11587.
[20] M. Alcaire, F. J. Aparicio, J. Obrero, C. López-Santos, F. J. Garcia-Garcia, J. R. Sánchez-Valencia, F. Frutos, K. (Ken) Ostrikov, A. Borrás, A. Barranco, *Advanced Functional Materials* **2019**, *29*, 1903535.
[21] A. Barranco, P. Groening, *Langmuir* **2006**, *22*, 6719.
[22] F. J. Aparicio, M. Holgado, A. Borras, I. Blaszczyk-Lezak, A. Griol, C. A. Barrios, R. Casquel, F. J. Sanza, H. Sohlström, M. Antelius, A. R. González-Elipe, A. Barranco, *Adv Mater* **2011**, *23*, 761.
[23] L. Contreras, J. Idígoras, A. Todinova, M. Salado, S. Kazim, S. Ahmad, J. A. Anta, *Physical Chemistry Chemical Physics* **2016**, *18*, 31033.
[24] L. Contreras-Bernal, M. Salado, A. Todinova, L. Calio, S. Ahmad, J. Idígoras, J. A. Anta, *The Journal of Physical Chemistry C* **2017**, *121*, 9705.
[25] L. Contreras-Bernal, S. Ramos-Terrón, A. Riquelme, P. P. Boix, J. Idígoras, I. Mora-Seró, J. A. Anta, *J. Mater. Chem. A* **2019**, *7*, 12191.
[26] A. Guerrero, G. Garcia-Belmonte, I. Mora-Sero, J. Bisquert, Y. S. Kang, T. J. Jacobsson, J.-P. Correa-Baena, A. Hagfeldt, *The Journal of Physical Chemistry C* **2016**, *120*, 8023.
[27] J.-P. Correa-Baena, M. Anaya, G. Lozano, W. Tress, K. Domanski, M. Saliba, T. Matsui, T. J. Jacobsson, M. E. Calvo, A. Abate, M. Grätzel, H. Míguez, A. Hagfeldt, *Advanced Materials* **2016**, *28*, 5031.
[28] E. J. Juarez-perez, M. Wußler, F. Fabregat-santiago, K. Lakus-wollny, E. Mankel, T. Mayer, W. Jaegermann, I. Mora-sero, *Role of the Selective Contacts in the Performance of Lead Halide Perovskite Solar Cells*.
[29] J.-P. Correa-Baena, S.-H. Turren-Cruz, W. Tress, A. Hagfeldt, C. Aranda, L. Shooshtari, J. Bisquert, A. Guerrero, *ACS Energy Letters* **2017**, *2*, 681.
[30] A. Castro-Chong, A. J. Riquelme, T. Aernouts, L. J. Bennett, G. Richardson, G. Oskam, J. A.

Anta, *ChemPlusChem* **2021**, *86*, 1347.
[31] S. R. Raga, E. M. Barea, F. Fabregat-Santiago, *The Journal of Physical Chemistry Letters* **2012**, *3*, 1629.
[32] J. Idígoras, G. Burdziński, J. Karolczak, J. Kubicki, G. Oskam, J. A. Anta, M. Ziółek, *The Journal of Physical Chemistry C* **2015**, *119*, 3931.
[33] L. Contreras-Bernal, A. Riquelme, J. J. Gallardo, J. Navas, J. Idígoras, J. A. Anta, *ACS Sustainable Chem. Eng.* **2020**, *8*, 7132.
[34] F. J. Aparicio, M. Alcaire, A. Borras, J. C. Gonzalez, F. López-Arbeloa, I. Blaszczyk-Lezak, A. R. González-Elipe, A. Barranco, *J. Mater. Chem. C* **2014**, *2*, 6561.

# Supporting Information

**Ultrathin plasma polymer passivation of perovskite solar cells for improved stability and reproducibility**

*Jose M. Obrero-Perez, Lidia Contreras-Bernal\*, Fernando Nuñez-Galvez, Javier Castillo-Seoane, Karen Valadez-Villalobos, Francisco J. Aparicio, Juan A. Anta, Ana Borras, Juan R. Sanchez-Valencia, and Angel Barranco\**

J. Obrero-Perez, Dr. L. Contreras-Bernal, F. Nuñez-Galvez, J. Castillo-Seoane, Dr. F. J. Aparicio, Dr. A. Borras, Dr. J.R Sanchez-Valencia and Dr. A. Barranco

Instituto de Ciencia de Materiales de Sevilla (CSIC-Universidad de Sevilla)

C/Americo Vespucio 49, E-41092 Seville, Spain

E-mail: lidia.contreras@icmse.csic.es; angel.barranco@csic.es

Dr. K. Valadez-Villalobos and J. A. Anta

Área de Química Física

Universidad Pablo de Olavide E-41013 Seville, Spain

**Table of contents**

**Experimental Section**

*Fabrication of Perovskite Solar Cells*

Glass coated of fluorine-doped tin oxide TEC 15 (FTO, Pilkington, resistance 15/square, 82-84.5 % transmittance) was used as the front electrode. Before use, FTOs were brushed with Hellmanex solution in water (2:98 vol %) and rinsed with deionized water. After that, the glasses were sequentially cleaned with Hellmanex solution, deionized water, isopropanol, and acetone using an ultrasonic bath for 15 min. The acetone was dried with a nitrogen flow. Then the substrates were treated with UV/O$_3$ for 15 min using an Ossila Ozone Cleaner. Once FTOs were clean, 30 nm of TiO$_2$ blocking layer was deposited on top by spray pyrolysis. For that, a precursor solution was prepared by addition of 1 mL of titanium diisopropoxide bis(acetylacetonate) solution (75% in 2-propanol, Sigma-Aldrich) in 14 mL of absolute ethanol. The solution was immediately sprayed on annealed glasses (450 °C) using oxygen as carrier gas. The substrates were kept at 450 °C for 30 min and then they were left to cool down to room temperature. This compact layer (c-TiO$_2$) was treated with UV/O$_3$ for 15 min before mesoporous TiO$_2$ deposition (*m*-TiO$_2$). The mesoporous solution was prepared by adding 1 mL absolute ethanol to 150 mg of a commercial TiO$_2$ paste (Sigma-Aldrich, 18NRT). The mesoporous dispersion was left under stirring overnight. After that, 100 μL were deposited on c-TiO$_2$ samples by spin-coating at 4000 rpm for 10 s. After spinning, the samples were immediately placed on a hot plate at 100 °C for 10 min. The *m*-TiO$_2$ layer was then sintered following a temperature programmer described by Saliba et. al.[8] The mesoporous layer once cooled was treated with UV/O$_3$ for 15 min as well. In the case of ADA (adamantane composite thin film) electrodes, the UV/O$_3$ treatment was 5 min. To deposit the perovskite film, the FTO/c-TiO$_2$/m-TiO$_2$, as well as FTO/c-TiO$_2$/m-TiO$_2$/ADA samples were transferred to a nitrogen glovebox (O$_2$ and H$_2$O levels below 0.5 ppm and temperature between 24 and 27 °C). The perovskite layer was prepared from a precursor perovskite solution consisting of mixing 1M formamidinium lead triiodide (FAPbI$_3$) and 1M methylammonium lead tribromide (MAPbBr$_3$) solutions (5/1%V, respectively) both in 1:4 %V DMSO:DMF (dimethyl sulfoxide and N,N-dimethylformamide, respectively) to which was added 5%V of 1.7M CsI solution in DMSO and them 5%V of RbI solution in 1:4 %V DMSO:DMF (0.2:99.8%mol, respectively).[8] The perovskite film (RbCsMAFA) was deposited by two-step spin-coating: 1) 1000 rpm, 10 s; 2) 6000 rpm, 20 s. 200 μL of clorobenzene was added as antisolvent in the second step 15 s after the beginning. After that, the samples were immediately placed on a hot plate at 100 °C for 60 min. As hole selective material, 70 mM 2,2,7,7-tetrakis[N,N-di(4-methoxyphenyl)amino]-9,9-spirobifluorene (Spiro-OMeTAD, Merck) in clorobenzene was used. This solution was doped with lithium bis(trifluoromethanesulfonyl)imide (LiTFSI, 520 mg/mL in acetonitrile), tris(2-(1H-pyrazol-1-yl)-4-tert-butylpyridine)- cobalt(III)tris(bis(trifluoromethylsulfonyl)imide) and 4-tert-Butylpyridine in a molar ratio of 0.5, 0.03 and 3.3, respectively. Finally, around 70 nm gold electrode were deposited by thermal evaporation in high vacuum.

*Deposition of adamantane on mesoporous TiO$_2$ electrode*

The adamantane plasma polymer thin film on top of *m*-TiO$_2$ samples was carried out in a microwave plasma reactor at base pressure of $10^{-6}$ torr as reported in previous works.[19–22] For this, we used an electron cyclotron resonance (ECR) plasma reactor with two separated zones for plasma and remote deposition. In the plasma zone, an argon microwave plasma (power 150 W and pressure $10^{-2}$ torr) was sustained and confined thanks to a set of magnets. The substrate holder was placed in a downstream configuration at a distance z = 9.5 cm from the glow discharge.[20] The samples were fixed to the backside of the holder where they were not directly exposed to the remote plasma discharge. Adamantane powder from Sigma-Aldrich was sublimated inside the chamber during the deposition from a heated container kept at 40 ºC. The film thickness and the deposition rate were monitored by using a quartz crystal microbalance (QCM) besides the sample holder.[34]

*Fabrication of dye sensitized solar cells*

Mesoporous TiO$_2$ electrodes covered with adamantane were used as working electrodes in made-to-purpose dye-sensitized solar cells (see the beginning of this section). The dye solution consisted of 0.3 mM N719 and 0.3 mM chanodeoxycholic acid in ethanol. As a counter electrode, a fluorine-doped tin oxide FTO8 (Pilkintong) coated with Platisol solution (Solaronix) was used. Both electrodes, working- and counter electrodes, were sealed together using a thin thermoplastic (Surlyn, Solaronix). The cavity was filled with an electrolyte composed of 0.03 M I2, 1 M 1-butyl-3-methylimidalozliumiodide, 0.05 M LiI, 0.5 M 4-terc-butylpiridine and 0.1 M Guanidine thiocynate in acetonitrile.[23]

*Characterization of films and devices*

Variable angle spectroscopic ellipsometry (VASE) was acquired in a Woollam V-VASE ellipsometer. The optical constants were modeling by fitting the spectra to the Cauchy model. Steady-state photoluminescence measurements were carried out using a Hitachi, F-7000. Steady state luminescence spectra were recorded in a Jobin Yvon Fluorolog-3 spectrofluorometer.

X-Ray Photoelectron Spectroscopy (XPS) characterization were performed in a Phoibos 100 DLD X-ray spectrometer from SPECS. The spectra were collected in the pass energy constant mode at a value of 50 eV using a Mg K$_\alpha$ source. Ti 2p3/2 signal at 458.5 eV was utilized for calibration of the binding energy in the spectra.

Current density-voltages (*J-V*) curves were recorded under a solar simulator (ABET-Sunlite) with AM 1.5G filter at 100 mW·cm$^{-2}$. these curves were measured with a black mask of 0.14 cm$^2$ to define the active area. For the case of ANOVA, the *J-V* curves were recorded at 50 mV s$^{-1}$ in reverse scan (from 1.2 V to -0.1 V) while for *HI* analysis the curves were measured at 60 mV s$^{-1}$. *HI* was calculated according to $HI = \frac{(J_{0.8V})_{reverse} - (J_{0.8V})_{forward}}{(J_{0.8V})_{reverse}}$. [23]

Scanning electron spectroscopy (SEM) images were obtained from Hitachi S4800 microscope operated at 2 kV.

Static wetting contact angles were determined by Youth method with a Data Physics Instruments using small droplets of 1 µL of pure-deionized water (Milli-Q) and the reported values are the average of seven measurements.

Impedance spectroscopy analysis (IS) was carried out using an analyzer module (PGSTAT302N/FRA2, Autolab) following our previously described methodology.[24,25] That is, the IS experiments were carried out at open circuit conditions by applying a 20 mV perturbation in the 106-0.1Hz. For this work, the illumination was provided by red (λ= 635 nm) light-emitting diode (LED) and over a wide range of DC light intensities. The stability study of perovskite solar cells devices was carried out under dark (test I) as well as under an indoor lighting cycle of 9 hours/days (test II). In both conditions, the samples were stored at ∼ 50% relative humidity and ∼ 23 ˚C. The *J-V* curves of these devices was measured under 1 sun – AM 1.5G illumination in the reverse scan using a scan rate of 60 mV·s$^{-1}$ in reverse scans. The active area was limited to 0.14 cm$^2$.

**Table S1.** Fitted thicknesses and optical constants for the Cauchy model of the adamantane nanocomposite films on mesoporous TiO$_2$ obtained by variable angle spectroscopic ellipsometry (VASE). The data in the table refer to ADA layers with UV ozone treatment.

| Device | Thickness [nm] | $A$n (550nm) | $B$n | $C$n |
|---|---|---|---|---|
| ADA-15nm | 15.3 ± 4.9 | 1.65 | 0.01 | 0 |
| ADA-5nm | 5.2 ± 1.9<br>6.2 ± 2.4* | 1.65 | 0.09 | 0 |
| ADA-1nm | 1.0 ± 0.1 | 1.65 | 0.3 | 0 |

*(No UV ozone treatment)

**Table S2.** Photovoltaic parameters statistic of perovskite solar cells with adamantane at mesoporous TiO$_2$/perovskite interface. Device without adamantane interlayer (ADA-0nm) was used as reference. The photovoltaic parameters were extracted from current-voltage curves measured under 1 sun-AM 1.5G illumination in reverse scan (from 1.2v to -0.1 V) using a mask of 0.14 cm$^2$. *HI* is the acronym for hysteresis index and it was obtained according to the equation from ref. [23] (for more details see experimental section).

| Device | $V_{OC}$ [V] | $J_{SC}$ [mA cm$^{-2}$] | FF [%] | PCE [%] | *HI* |
|---|---|---|---|---|---|
| ADA-0nm | 1.07 ± 0.04 | 18.9 ± 1.3 | 74 ± 4 | 14.9 ± 1.5 | 0.13 ± 0.07 |
| ADA-1nm | 1.08 ± 0.02 | 19.6 ± 0.7 | 74 ± 2 | 15.7 ± 0.9 | 0.13 ± 0.02 |
| ADA-5nm | 1.12 ± 0.02 | 19.7 ± 0.6 | 77 ± 1 | 16.8 ± 0.8 | 0.05 ± 0.02 |
| ADA-15nm | 1.09 ±0.02 | 20.0 ± 1.1 | 74 ± 3 | 16.3 ± 0.9 | 0.08 ± 0.01 |

**Table S3.** *β* parameter values extracted from the high frequency resistance fitted from Nyquist spectra for ADA and non-ADA samples. These spectra were recorded at open circuit potential and using red light (LED) as illumination source.

| Sample | Slope (β) |
|---|---|
| ADA-15nm | 1.6 |
| ADA-5nm | 2.2 |
| ADA-1nm | 1.6 |
| ADA-0nm | 2.3 |

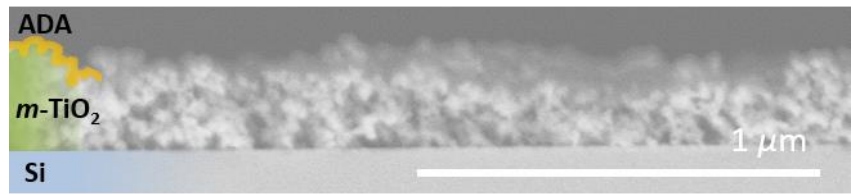

**Figure S1.** Cross-sectional SEM image of 15 nm adamantane plasma polymer on top of mesoporous TiO$_2$ layer.

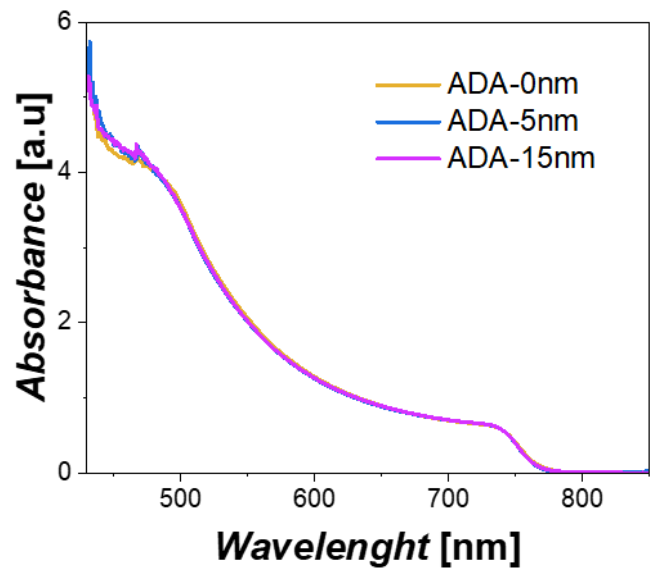

**Figure S2.** UV-Vis spectra of perovskite films deposited on the mesoporous TiO$_2$ electrodes: with (ADA-5nm and -15nm) and without ADA (ADA-0nm) polymer at *m*-TiO$_2$/perovskite interface.

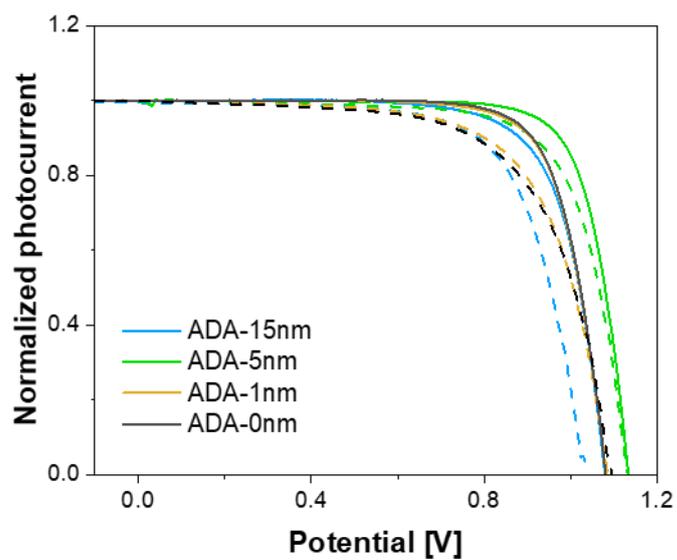

**Figure S3.** Normalized (reference voltage: -0.1 V) current density–voltage curves measured in the reverse scan (line) and forward scan (dash line) at 60 mv s$^{-1}$ under 1 sun-AM 1.5G illumination.